\RequirePackage{fix-cm}
\documentclass[smallextended]{svjour3}       
\smartqed  
\usepackage{graphicx}
\usepackage{mathptmx}      
\usepackage{amssymb}
\begin{document}

\title{Cosmological black holes and the direction of time}

\author{Gustavo E. Romero, Daniela P\'erez, and Federico G. Lopez Armengol}



\institute{Instituto Argentino de Radioastronom{\'{i}}a (IAR, CCT La Plata, CONICET) \at
              C.C. No. 5, 1894, Villa Elisa, Buenos Aires, Argentina. \\
              Tel.: +54-221-482-4903\\
              Fax: +54-221-425-4909\\
              \email{romero@iar-conicet.gov.ar}
}

\date{Received: date / Accepted: date}




\date{Received: date / Accepted: date}

\maketitle

\begin{abstract}
Macroscopic irreversible processes emerge from fundamental physical laws of reversible character. The source of the local irreversibility seems to be not in the laws themselves but in the initial and boundary conditions of the equations that represent the laws. In this work we propose that the screening of currents by black hole event horizons determines, locally, a preferred direction for the flux of electromagnetic energy. We study the growth of black hole event horizons due to the cosmological expansion and accretion of cosmic microwave background radiation, for different cosmological models. We propose generalized McVittie co-moving metrics and integrate the rate of accretion of cosmic microwave background radiation onto a supermassive black hole over cosmic time. We find that for flat, open, and closed Friedmann cosmological models, the ratio of the total area of the black hole event horizons with respect to the area of a radial co-moving space-like hypersurface always increases. Since accretion of cosmic radiation sets an absolute lower limit to the total matter accreted by black holes, this implies that the causal past and future are not mirror symmetric for any spacetime event. The asymmetry causes a net Poynting flux in the global future direction; the latter is in turn related to the ever increasing thermodynamic entropy. Thus, we expose a connection between four different ``time arrows'': cosmological, electromagnetic, gravitational, and thermodynamic.
\end{abstract}


\section{Introduction}
\label{sec:intro}

Physical laws are regular patterns in the occurrence of events. Events, in turn, are changes in the properties of things (Bunge1967). It is a remarkable fact that all formal representations of the fundamental laws of physics are invariant under the operation of time reversal, whereas macroscopic processes are not symmetric with respect to time. There seems to be a preferred direction for the occurrence of events, the so-called arrow of time\footnote{The expression ``arrow of time'' is misleading since time is not a vector. Time is an emergent property of changing things and can be represented by a one-dimensional continuum. Time itself is not asymmetric; the processes time parametrizes may be asymmetric.}. Invariably, the ultimate fate of all things is a state of equilibrium, of maximum entropy as expressed in the Second Law of Thermodynamics. How is possible that irreversible macroscopic processes can emerge in a universe where all fundamental physical laws are of reversible character?

As pointed out by Gold (1962) and Penrose (1979), the origin of the irreversibility is not in the laws but in the initial and boundary conditions under which the laws operate. But how is possible for conditions that are of global nature to enforce local proceses as those represented by the laws of thermodynamics?

The physical interactions that dominate on medium (human) scales are those of electromagnetic origin. The strong and weak interactions are of very short range. Gravity, instead, rules on cosmological scales; it is extremely weak in comparison with electromagnetism on short ranges. If there is a global-to-local connection that enforces irreversibility, it should 
manifest as some kind of coupling between gravitational and electromagnetic interactions. 

The electromagnetic radiation field can be described in terms of a four-potential $A^{\mu}$, which satisfies Maxwell's equations in the Lorentz gauge:
\begin{equation}\label{maxwell}
\partial^{\nu} \partial_{\nu} A^{\mu}(\textbf{r},t)=4 \pi j^{\mu}(\textbf{r},t),
\end{equation}
where we take $c = 1$ and $j^{\mu}$ denotes the four-current.

The solution $A^{\mu}$ is a functional of the currents $j^{\mu}$. Equation (\ref{maxwell}) admits both retarded and advanced solutions: 

\begin{equation}\label{ret}
A^{\mu}_{\rm ret}(\vec{r},\;t)=\int_{V_{\rm ret}}
\frac{j^{\mu} \left(\vec{r},\;t-\left|\vec{r}-\vec{r'}\right|\right)}{\left|\vec{r}-\vec{r'}\right|}d^{3}\vec{r'}, 
\end{equation}

\begin{equation}\label{adv}
A^{\mu}_{\rm adv}(\vec{r},\;t)=\int_{V_{\rm adv}}
\frac{j^{\mu} \left(\vec{r},\;t+\left|\vec{r}-\vec{r'}\right|\right)}{\left|\vec{r}-\vec{r'}\right|}d^{3}\vec{r'}. 
\end{equation}
The two functionals of $j^{\mu}(\vec{r}, t)$ are related to one another by a time reversal transformation. Solution (\ref{ret}) is contributed by currents in the causal past ($V_{\rm ret} = J^{-}(p)$) of the spacetime point $p(\vec{r}, t)$, and solution (\ref{adv}) by currents in the causal future ($V_{\rm adv} = J^{+}(p)$) of that point. We assume here Sommerfeld radiation condition, that makes source-free radiation null, and hence the total border contribution to the solutions is zero.

The linear combinations of electromagnetic solutions are also solutions, since the equations are linear and the Principle of Superposition holds. It is
usual to consider only the retarded potential as physical meaningful in order to estimate the electromagnetic field at any arbitrary point  $p(\vec{r}, t)$: $F^{\mu\nu}_{\rm ret}=\partial^{\mu} A^{\nu}_{\rm ret}- \partial^{\nu}A^{\mu}_{\rm ret}$.  However, this is  an {\it ad hoc} assumption that breaks time symmetry, introducing a time-oriented causality principle. There seems to be no compelling reason for such a choice, which actually is tantamount to postulate a privileged time direction giving up any attempt at a deeper explanation\footnote{See Sciama (1967) and especially Clarke (1977). See also the ``Wheeler-Feynmann absorber theory'' (Wheeler and Feynman 1945, 1949) .}. We can adopt, for instance, a solution of the form:
\begin{equation}\label{suma}
A^{\mu}(\vec{r}, t)=\frac{1}{2}\left(\int_{J^{-}} {\rm ret} +\int_{J^{+}} {\rm adv}\right) dV,
\end{equation}
which is formally valid and treats equally the causal past and future of the event at $(\vec{r}, t)$.

In Minkowski spacetime, the light cone that determines the local causal structure is mirror symmetric with respect to a space-like surface that contains the point $p(\vec{r},t)$. Hence, the averaged sources in the causal past and causal future are the same, the boundary conditions are the same, and the retarded and advanced solutions are identical. This is not necessarily the case in the presence of gravitation: because of the spacetime curvature, the light cones ought  not be mirror symmetric with respect to the space-like hyperplane that includes the point $p(\vec{r},t)$. Hence, the contributions of the past and future currents might differ.


Romero and P\'erez (2011) recently showed that if the universe is accelerating (Perlmuter et al. 1999), then $J^{-}(p)$ and $J^{+}(p)$ are not mirror-symmetric because of the presence of particle cosmological horizons. This, in turn, implies that $A^{\mu}_{\rm ret}$ and $A^{\mu}_{\rm adv}$ will be different. From Eq. (\ref{suma}), a vector field $L^{\mu}$ can be defined as:
\begin{equation}\label{asy}
L^{\mu} = \left(\int_{J^{-}} {\rm ret} -\int_{J^{+}} {\rm adv}\right) dV \neq 0.
\end{equation} 
If $g_{\mu \nu}L^{\mu}T^{\nu} \neq 0$ where $T^{\nu} = (1,0,0,0)$, there is a preferred direction for the flux of electromagnetic energy in spacetime towards the global future direction given by the gradient of the currents distribution (Romero and P\'erez 2011).

In cosmological models without accelerated expansion, the asymmetry described by Romero and P\'erez (2011) does not hold because of the absence of particle horizons. We propose now that black holes might also cause an asymmetry in the contribution of currents in the causal past with respect to the currents in the causal future of any event. This allows us to define a ``time direction'' even in universes that are not in a state of accelerated expansion (e.g. Friedmann cosmological models).

A black hole is a region from where no null
geodesic can reach the asymptotic flat future space-time (for a full characterization see, e.g. Wald 1984). These objects
are spacetime regions that are causally disconnected from the rest of the universe. The boundary surface between the interior of the black hole and the exterior is called event horizon. Events enclosed by the horizon cannot make any influence on the outside world. In particular, currents inside black holes have no effect outside. 


The aim of this work is to investigate the effect of the screening of currents by event horizons in the solutions of Maxwell's equations that locally determine the electromagnetic flux. In particular, we show that the ratio of the area of the event horizon with respect to the area of a radial space-like hypersurface in Friedmann cosmological models, always increases. We consider that the minimum growth of event horizons is caused by the cosmological expansion and the accretion of radiation of the cosmic microwave background (CMB) to which all black holes in the universe are exposed. This sets a lower limit to the growth of black holes. The result is that larger black holes in the future will hide more currents than black holes in the past of any event, with the consequent asymmetry resulting from Eq. (\ref{asy}).

Our paper is organized as follows: in Section \ref{Friedmanncosmomodels} we describe the Friedmann cosmological models. In Section \ref{growthbh} we study the two processes mentioned above that contribute to the growth of black holes in a cosmological context. In the next section we define a black hole filling factor function. Section \ref{ff-cosmo-models} is devoted to the calculation of this filling factor for the Friedmann cosmological models. Finally, a discussion of the results obtained is offered in Section \ref{conclusions}.The more technical details are confined to the Appendix. 

\section{Friedmann cosmological models}
\label{Friedmanncosmomodels}

Friedmann cosmological models are particular cases of Friedmann-Lema\^itre-Robertson-Walker (FLRW) models. They are characterized by a zero cosmological constant and a negligible ratio of radiation to matter energy density. The spacetime line element of these models in co-moving coordinates $(t,r,\theta,\phi)$ is the well-known FLRW interval:
\begin{eqnarray}
ds^{2} & = & - c^{2} dt^{2} + {a(t)}^{2}\frac{dr^{2}}{1-k r^2 R_0^{-2}}\nonumber \\
& + & {a(t)}^{2} r^{2}\left({d\theta}^{2}+{\sin{\theta}}^{2} {d\phi}^{2}\right).\label{FLRW}
\end{eqnarray}
Here $a(t)\equiv R(t)/R_0$ is the normalized scale factor, where $R_0$ is the scale factor at the present epoch, and $k$ takes the values $-1$, $0$, or $1$ depending on whether the spatial section has negative, zero, or positive curvature, respectively.

The normalized scale factor $a(t)$ depends crucially on the sign of the spatial curvature. If $\Omega_{\rm m,0}$ and $\Omega_{\rm k,0}$ denote the present\footnote{By `present' we mean the values of such parameters if the universe would be today matter dominated.} normalized matter and curvature energy densities, respectively, the normalized scale factors for the closed, flat and open Friedmann models are given by:
\begin{itemize}
\item Closed Friedmann model ($k>0$):
\begin{eqnarray}
\label{closed}
a & = &\frac{\Omega_{\rm m,0}}{2(\Omega_{\rm m,0}-1)}(1-\cos \varphi), \label{eliptica-a}\\
t & = &\frac{\Omega_{\rm m,0}}{2H_0 (\Omega_{\rm m,0}-1)^{3/2}}(\varphi - \sin \varphi),\label{eliptica-t}
\end{eqnarray}
where the parameter $\varphi\in[0;2\pi]$.\\
			
\item Flat Friedmann model ($k=0$):
			\begin{equation}
			\label{parabolica}
			a(t)=\left(\frac{3}{2}H_0 t \right)^{2/3}.
			\end{equation}

\item Open Friedmann model ($k<0$):
			\begin{eqnarray}
			\label{open}
				a & = &\frac{\Omega_{\rm m,0}}{2(1-\Omega_{\rm m,0})} (\cosh \varphi -1), \label{hiperbolica-a}\\
				t & = &\frac{\Omega_{\rm m,0}}{2H_0 (1-\Omega_{\rm m,0})^{3/2}} (\sinh \varphi -\varphi),\label{hiperbolica-t}
			\end{eqnarray}
			where the parameter $\varphi>0$.
\end{itemize}

We adopt for the Hubble constant the value recently obtained by the Planck Collaboration (2014). :
\begin{equation}
H_{0} = 67.3 \frac{{\rm km}}{{\rm s\:Mpc}}.
\end{equation}

The values of the cosmological parameters\footnote{The values of the matter and curvature densities in the present epoch were calculated using the equation:
\begin{equation}
\Omega_{\rm m,0}+\Omega_{\rm k,0}=1.
\end{equation}
The values of the present scale factor were obtained from:
\begin{equation}
R_{0} = c {\int_{0}}^{t_{0}} \frac{dt}{a(t)}.
\end{equation}} $\Omega_{\rm k,0}$, $\Omega_{\rm m,0}$, $R_{0}$, and $t_{0}$ used in this work are given in Table \ref{tab:1}.

\begin{table}[tbp]
\centering
\begin{tabular}{|c|ccc|}
\hline
& Closed FCM & Flat FCM & Open FCM  \\
\hline
$\Omega_{\rm k,0}$ & $-0.30$ & $0$ & $0.21$ \\
$\Omega_{\rm m,0}$ & $1.30$ & $1$ & $0.79$ \\
$R_{0}$  $\left[{\rm cm}\right]$& $2.52 \times 10^{28}$ & $2.75 \times 10^{28}$ & $2.98 \times 10^{28}$\\
$t_{0}$  $\left[{\rm s}\right]$& $2.90 \times 10^{17}$ & $3.06 \times 10^{17}$ & $3.20 \times 10^{17}$\\ 
\hline
\end{tabular} 
\caption{Cosmological parameters for closed, flat, and open Friedmann cosmological models: $\Omega_{\rm k,0}$, $\Omega_{\rm m,0}$, $R_0$, and $t_0$  stand for the values of the curvature density, the matter density, the scale factor at $t_0$, and the age of the universe, respectively.}
\label{tab:1}
\end{table}


\section{Growth of black holes in cosmological contexts}
\label{growthbh}

Two distinctive processes at least contribute to the growth of a black hole along cosmic time $t$. First, black holes do participate in the cosmological expansion; since black holes are essentially spacetime regions, event horizons increase their size along with the whole spacetime (Nandra et al. 2012a,b; Faraoni 2015). We propose generalized McVittie co-moving metrics that are in agreement with this hypothesis. Second,  black holes unavoidably accrete photons from the CMB. This imposes an absolute lower limit to the mass increase by accretion. Our estimation is approximate because we take radiation into account in our treatment of the local physics but neglect it on global cosmological dynamics. The approximation, however, is a very good one for most of the history of any universe. A detailed account of both processes is given in the Appendix.

\section{Filling factor}
\label{fillingfactor}

We define the {\it filling factor}, denoted $f(t)$, as the ratio between the area of the black hole event horizon $A_{\footnotesize{\rm BH}}(t) $ and the area of a radial spacelike hypersurface in the Friedmann models $\Sigma(t)$:
\begin{equation}\label{ff}
f(t) \equiv \frac{A_{\footnotesize{\rm BH}}(t)}{\Sigma (t)}. 
\end{equation}

The area of the black hole event horizon as a function of cosmic time $t$ takes the form:
\begin{equation}\label{eh}
A_{\footnotesize{\rm BH}}(t) = \frac{16 \pi G^{2}}{c^{4}} {{ M_{\footnotesize{\rm BH}}^{\footnotesize{\rm TOT}}}(t)}^{2},
\end{equation}
where
\begin{equation}
{ M_{\footnotesize{\rm BH}}^{\footnotesize{\rm TOT}}}(t) =  M_{\footnotesize{\rm BH}}^{\footnotesize{\rm HH}}(t) + \Delta M_{\footnotesize{\rm BH}}^{\footnotesize{\rm CMB}}(t).
\end{equation}

The area of a radial spacelike hypersurface in the Friedmann models can be derived from the line element (\ref{FLRW}) and turns to be:
\begin{equation}\label{hyper-s}
\Sigma(t) = 4 \pi {R_{0}}^{2} r^{2} a(t)^{2}.
\end{equation}

Given the expressions above, we get a formula for $f(t)$ from Eqs. (\ref{ff}), (\ref{eh}), and (\ref{hyper-s}):
\begin{equation}\label{ff1}
f(t) = \frac{4 \:G^{2}}{c^{4}\: R_{\rm 0}^{2}\: r^{2}\: {a(t)}^{2}}\left[M_{\rm 0} a(t)+\Delta M_{\footnotesize{\rm BH}}^{\footnotesize{\rm CMB}}(t)\right]^{2} .
\end{equation}
We impose the condition that the value of the filling factor must be equal to $1$ at the present cosmic time $t_{0}$. Then, we take the radial hypersurface with co-moving coordinate:
\begin{equation}
r = \frac{2GM_{0}}{c^{2}R_{0}}.
\end{equation}
The filling factor now takes the form:
\begin{equation}\label{final-f}
f(t)=1+\frac{2 \Delta M_{\footnotesize{\rm BH}}^{\footnotesize{\rm CMB}}(t)}{M_0 a(t)}+ \frac{\Delta M_{\footnotesize{\rm BH}}^{\footnotesize{\rm CMB}}(t)^2}{M_0^2a(t)^2}.
\end{equation}

In what follows we shall prove that the function $f(t)$ is greater than its present value $f(t_0)=1$, for $t>t_0$ in the open, flat, and closed Friedmann cosmological models. This will result in a time asymmetry in these models. 


\section{Results: filling factor for Friedmann cosmological models}
\label{ff-cosmo-models}

\begin{figure}[tbp]
\centering
\includegraphics[width=.9\columnwidth]{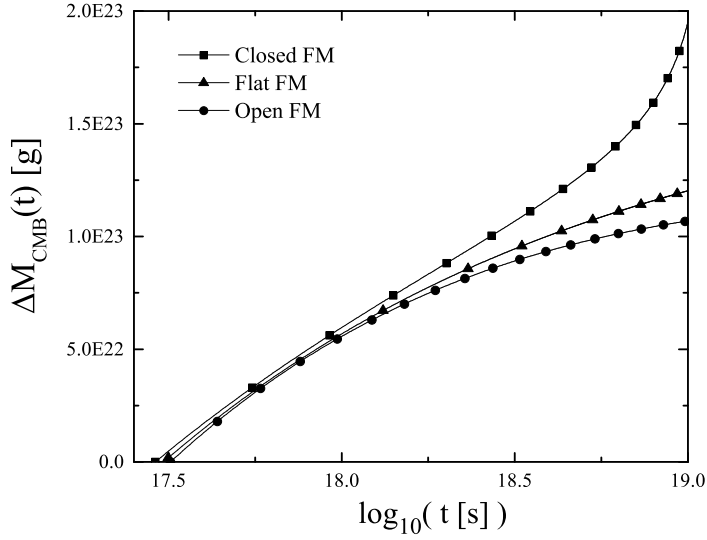}
\caption{Plots of the function $\Delta M_{\footnotesize{\rm BH}}^{\footnotesize{\rm CMB}}(t)$ as a function of the cosmic time for the closed, flat, and open Friedmann models (FM)}
\label{Mcmbfigure}
\end{figure}

We show in Figure \ref{Mcmbfigure} plots of Eq. (\ref{masa-CMB}) as a function of the cosmic time $t$ for the closed, flat, and open Friedmann models. We see that the total mass of the black hole always increases due to the accretion of CMB photons. In the flat and open models, $\Delta M_{\footnotesize{\rm BH}}^{\footnotesize{\rm CMB}}(t)$ approximately tends to a constant value as $t\rightarrow \infty$. The density and energy of the CMB photons is inversely proportional to the normalized scale factor (Eqs. (\ref{density-ph}) and (\ref{energy-ph})). For a universe that always expands, such as the flat or open models, black holes accrete less and less energetic photons as $t \rightarrow \infty$. Contrary, in the closed case, as the universe collapses, $\Delta M_{\footnotesize{\rm BH}}^{\footnotesize{\rm CMB}}(t)$ tends to infinity.

\begin{figure}[tbp]
\centering
\includegraphics[width=.9\columnwidth]{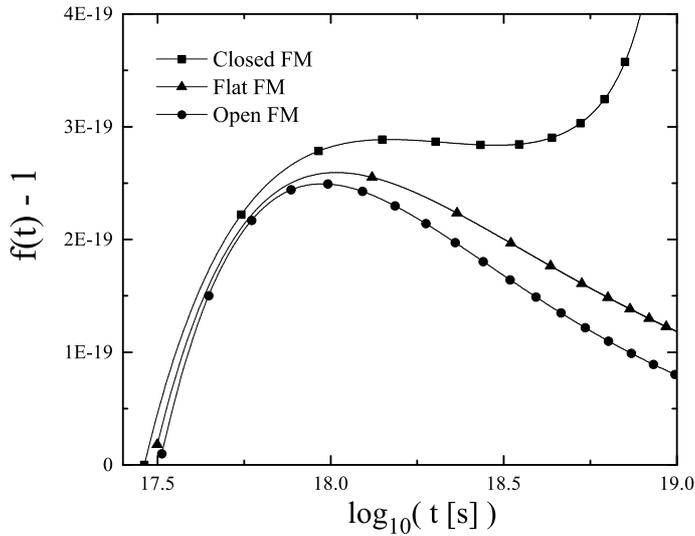}
\caption{Plots of the function $f(t)-1$ as a function of the cosmic time for the closed, flat, and open Friedmann models (FM).} 
\label{fm1figure}
\end{figure}

In Figure \ref{fm1figure} we plot the function $f(t)-1$ (see Eq. (\ref{final-f})) as a function of the cosmic time. In the three cosmological models, $f(t)-1$ is always positive. For both the flat and open models, $f(t)-1$ increases until a maximum value at $t \approx 10^{18} {\rm s}$, and it tends to zero for higher values of the cosmic time. This is not the case for the closed model; the function $f(t)-1$ has a local maximum and a local minimum at $t \approx 10^{18.2} {\rm s}$ and $t \approx 10^{18.5} {\rm s} $, respectively. For $t \rightarrow \infty$, $f(t)-1$ goes to infinity. 

\section{Implications and conclusions}
\label{conclusions}

We have shown that the filling factor $f(t)$ is always larger than its present value $f(t_0)=1$, for $ t > t_{0}$ in the flat, open, and closed Friedmann cosmological models. The growth of the area of the black hole event horizons is always greater than the growth of the area of a space-like hypersurface in cosmological expansion. The causal structure of a given point of spacetime $p(\textbf{r},t)$ is determined by the causal past and future of $p(\textbf{r},t)$. There is an asymmetry in the contributions of currents in the causal past with respect to the causal future due to the screening of currents of larger black hole event horizons in the future. This implies that $J^{-}(p)$ and $J^{+}(p)$ are not mirror symmetric. Consequently, the four-potentials $A^{\mu}_{{\rm ret}}$ and $A^{\mu}_{{\rm adv}}$ are different. In all cases this implies that $g_{\mu \nu}L^{\mu}T^{\nu} > 0$.

Even for a contracting universe, such as the closed Friedmann model, macroscopic processes are still irreversible. This is contrary to Hawking's prediction Hawking (1985) that the direction in which processes occur should reverse in a recollapsing phase of the universe.  

It might be argued that in flat and open models there will be a time when the temperature of the black holes will be higher than the temperature of the CMB and then black hole evaporation will become effective. This is correct, but has no effect on our result since Hawking radiation is thermal (Hawking 1974, 1975) and the screened currents can never be recovered. This leaves the time asymmetry untouched even when quantum field theory is applied to the horizons. Our treatment here, however, is classical. For more discussion on the so-called ``information paradox'' of black holes, we refer the reader to the recent review by Romero (2014).

Penrose (Penrose 1979) distinguished several asymmetric fundamental processes in the universe which seem to be independent: 1) the Second Law of Thermodynamics, 2) the arrow of radiation, 3) the expansion of the universe, 4) the formation of black holes\footnote{The Second Law of Thermodynamics: the entropy of a closed system never decreases. The arrow of radiation: we only measure retarded potentials, though retarded and advances potentials are both solutions of Maxwell's equations. The expansion of the universe: there are several asymmetric processes involved in the expansion of the universe such as the receding of distant galaxies or the diminution of the mean universal temperature. Black hole formation: black and white holes are both solutions of Einstein's equations, related by a time reversal transformation. However, only black holes seem to exist.}. Our work implies that the four asymmetric processes mentioned above are all related in the Friedmann models. As we have shown, there is a preferred temporal direction for the flux of electromagnetic energy towards global future. The electromagnetic flux $L$ is related to the macroscopic temperature through the Stefan-Boltzmann law: $L = A \sigma_{{\rm SB}} T^{4}$, where $\sigma_{{\rm SB}}$ is the Stefan-Boltzmann constant. If we assume local energy conservation, the temperature's body will decrease towards the future, increasing its entropy according to the Second Law of Thermodynamics. The arrow of radiation and the Second Law of Thermodynamics emerge from the screening of currents by larger black holes as cosmic time increases. 

We conclude that the coupling between global cosmological conditions and gravitational processes with local electromagnetic events gives rise to the local macroscopic irreversibility observed in the universe. This irreversibility would exist even in the case of a recollapsing universe with black holes.   

\begin{acknowledgements}
This work is supported by grant PIP 0338,(CONICET) and grant AYA2016-76012-C3-1-P (Ministro de Educaci\'on, Cultura y Deporte,
Espa\~na).
\end{acknowledgements}

\section*{Appendix: Technical details of the cosmological growth of black holes.}
\label{Appendix}

\subsection*{A.1  Cosmological expansion}

Black holes are co-moving with the cosmological expansion of the universe. In order to describe the increase of size of the black hole due to this expansion, we adopt the Hawking-Hayward quasilocal mass (Hawking 1968, Hayward 1994), which measures the mass of a bound source of gravitation in an asymptotically FRWL universe (Nolan 1998). Specifically, we adopt:
\begin{equation}\label{massHH}
M_{\footnotesize{\rm BH}}^{\footnotesize{\rm HH}}(t) = M_{\rm 0}\: a(t),
\end{equation}
where $M_{\rm{0}}$ is the black hole present mass and $a(t)$ stands for the normalized scale factor. To implement this proposal, we replace Hawking-Hayward quasi-local mass in McVittie metrics (McVittie 1933) for a mass-particle embedded in a Friedmann cosmological model and, via Einstein field equations, we study the corresponding energy-momentum tensor.

For a cosmological black hole embedded in the flat Friedmann model, the generalized McVittie metric in isotropic coordinates $(t,r,\theta,\phi)$ is (Gao et al. 2008):
\begin{eqnarray}
ds^2 & = & -\frac{\left\lbrace1-\frac{M(t)}{2\:r a(t)}\right\rbrace^2}{\left\lbrace1+\frac{M(t)}{2\:ra\:(t)}\right\rbrace^2} dt^2\nonumber \\ 
& + &  a(t)^2 \left\lbrace1+\frac{M(t)}{2\:r\:a(t)}\right\rbrace^4 \left(dr^2+r^2d\Omega^2\right),\label{mcvittieflat}
\end{eqnarray}
where $M(t)$ is an arbitrary function of the cosmic time $t$. For the open and closed Friedmann models, we propose the following metrics:

\begin{eqnarray}\label{mcvittiecurved}
ds^2= & - &\frac{\left\lbrace1-\frac{M(t)}{2\:r\:a(t)} \left(1\pm \frac{r^2}{4R_0^2}\right)^{1/2}\right\rbrace^2}{\left\lbrace1+\frac{M(t)}{2\:r\:a(t)} \left(1 \pm \frac{r^2}{4R_0^2}\right)^{1/2}\right\rbrace^2} dt^2 \label{mcvittiecurved} \\
& + &\frac{\left\lbrace1+\frac{M(t)}{2\:r\:a(t)} \left(1 \pm \frac{r^2}{4R_0^2}\right)^{1/2}\right\rbrace^4}{\left(1 \pm \frac{r^2}{4R_0^2}\right)^2} a(t)^2 dr^2\nonumber \\
& + &\frac{\left\lbrace1+\frac{M(t)}{2\:r\:a(t)} \left(1 \pm \frac{r^2}{4R_0^2}\right)^{1/2}\right \rbrace^4 a(t)^2 r^2 d\Omega^2}{\left(1 \pm \frac{r^2}{4R_0^2}\right)^2}, \nonumber
\end{eqnarray}
where $\pm$ is the sign of the spatial curvature. If we adopt for the function $M(t)$ the Hawking-Hayward quasi-local mass, given by Eq. \ref{massHH}, the  metrics above take the form:
\begin{eqnarray}
ds^2 & = &-\frac{\left\lbrace1-\frac{M_0}{2\:r}\right\rbrace^2}{\left\lbrace1+\frac{M_0}{2\:r}\right\rbrace^2} dt^2\nonumber \\
& + & a(t)^2 \left\lbrace1+\frac{M_0}{2\:r}\right\rbrace^4 \left(dr^2+r^2d\Omega^2\right),\\
ds^2= & - &\frac{\left\lbrace1-\frac{M_0}{2\:r} \left(1\pm \frac{r^2}{4R_0^2}\right)^{1/2}\right\rbrace^2}{\left\lbrace1+\frac{M_0}{2\:r} \left(1 \pm \frac{r^2}{4R_0^2}\right)^{1/2}\right\rbrace^2} dt^2 \nonumber \\
& + &\frac{\left\lbrace1+\frac{M_0}{2\:r} \left(1 \pm \frac{r^2}{4R_0^2}\right)^{1/2}\right\rbrace^4}{\left(1 \pm \frac{r^2}{4R_0^2}\right)^2} a(t)^2 dr^2 \nonumber \\
& + & \frac{\left\lbrace1+\frac{M_0}{2\:r} \left(1 \pm \frac{r^2}{4R_0^2}\right)^{1/2}\right\rbrace^4}{\left(1 \pm \frac{r^2}{4R_0^2}\right)^2} a(t)^2   r^2 d\Omega^2,
\end{eqnarray}
Notice that if $a(t)= 1$ and $r<<R_0$, the metrics (\ref{mcvittieflat}) and (\ref{mcvittiecurved}) reduce to the Schwarzschild metric in physical isotropic coordinates $(t,a(t)r,\theta,\phi)$ with central mass $M_0 a(t)$. Conversely, if $M_0=0$, we recover the Friedmann geometries.  

The asymptotically flat metric (\ref{mcvittieflat}) and its corresponding energy-momentum tensor has been analyzed by Gao et al. (2008). The metric turns to be consistent with the energy-momentum of an imperfect pressureless accreting fluid. We emphasize that, asymptotically, the energy-momentum tensor coincides with the one of the cosmological model:
\begin{equation}
T_{\mu \nu}=\rho(r,t) u_\mu u_\nu + q_\mu u_\nu + q_\nu u_\mu \rightarrow
\; \rho_\infty(t) u_\mu u_\nu.
\end{equation}
Here $u^\mu \rightarrow (|g_{00}(r,t)|^{-1/2},0,0,0)$, $q^\mu \rightarrow (0,q(r,t),0,0)$, and $\rho_\infty\propto H^2(t)$, in agreement with Friedmann equations\footnote{Friedmann equation:\begin{equation} H^2(t)\equiv\left(\frac{\dot{a}(t)}{a(t)}\right)^2=\frac{8\pi G}{3} \rho_m(t) - \frac{c^2 k}{R_0^2 a^2(t)}.\end{equation}}.

We maintain that the same result can be extend to the asymptotically curved metrics (\ref{mcvittiecurved}). According to Einstein field equations, the matter content of these geometries is an imperfect accreting fluid that, asymptotically, coincides with the one of curved Friedmann models:
\begin{eqnarray}
T_{\mu \nu} & = & [P(r,t)+\rho(r,t)] u_\mu u_\nu \nonumber\\
& + &  P(r,t) g_{\mu \nu} + q_\mu u_\nu + q_\nu u_\mu \nonumber \\ 
& &\rightarrow
\; \rho_\infty(t) u_\mu u_\nu, \nonumber
\end{eqnarray}
where $\rho_\infty(t) \propto H^2(t) \pm C a^{-2}(t)$, with $C$ a positive constant, in accordance with Friedmann equations.

We have shown that the spacetime metrics given by (\ref{mcvittieflat}) and (\ref{mcvittiecurved}) are consistent with Hawking-Hayward quasi-local mass proposal for a black hole embedded in Friedmann cosmological models. A detailed analysis of the existence of trapped surfaces for those metrics will be given in a future work. In what follows, we will assume Hawking-Hayward quasi-local mass as the effective mass of a black hole embedded in a Friedmann cosmological model.

\subsection{A.2  Black hole accretion of cosmic microwave background radiation}

The rate of accretion of CMB photons onto a black hole is given by:
\begin{equation}\label{accretion-CMB}
\dot{M}_{\footnotesize{\rm BH}}^{\footnotesize{\rm CMB}} = \pi b^{2} n_{\footnotesize{\rm CMB}}(z) <{\epsilon}_{{\rm ph}}> c,
\end{equation}
where $b = \left(3 \sqrt{3}r_{\rm S}\right)/2 $ is the critical impact parameter of a Schwarzschild black hole ($r_{\rm S}$ stands for one Schwarzschild radius); $n_{\footnotesize{\rm CMB}}(z)$ and $ <{\epsilon}_{{\rm ph}}>$ are the density and energy of CMB photons as a function of redshift $z$, respectively:
\begin{eqnarray}\label{density-ph}
n_{\footnotesize{\rm CMB}}(z) & = & 5\:10^{-2} {\rm cm}^{-3} \left(1+z\right)^{3},\\
 <{\epsilon}_{{\rm ph}}> & = & 6\:10^{-6} {\rm erg} \left(1+z\right).\label{energy-ph}
\end{eqnarray} 

For simplicity, we consider only supermassive black holes\footnote{Supermassive black holes are found in most galaxies and have masses from $\approx 10^{6}$ to $10^{9}M_{\odot}$ (Romero and Vila 2013). The contribution to the screening of currents by stellar mass black holes, in comparison, is negligible.} of $M_{0} = 10^{8}\:M_{\odot}$. Equation (\ref{accretion-CMB}) and (\ref{massHH}) then yields:
\begin{equation}
\dot{M}_{\footnotesize{\rm BH}}^{\footnotesize{\rm CMB}} (t)= 5.5\:10^{25}\:\pi \left(\frac{a_{\rm 0}}{a(t)}\right)^{2}\: \frac{{\rm erg}}{{\rm s}}.
\end{equation}  

The variation of black hole mass by accretion of CMB photons during the cosmic time interval $\Delta t = t-t_{0}$ ($t_{0}$ is the value of the present cosmic time) can be obtained by integration of the latter equation:
\begin{equation}\label{masa-CMB}
\Delta M_{\footnotesize{\rm BH}}^{\footnotesize{\rm CMB}}(t) = {\int_{t_{\rm 0}}}^{t} \frac{\dot{M}_{\footnotesize{\rm BH}}^{\footnotesize{\rm CMB}} (t')}{c^{2}} dt'.
\end{equation}

As mentioned above, it is expected that accretion of matter and currents also occurs, as it is actually observed in active galactic nuclei (Romero and Vila 2013). In this sense, our estimation for the growth of the black hole mass due to CMB accretion gives the absolute minimum.





\newpage

\section*{Gustavo E. Romero} Full Professor of Relativistic Astrophysics at the University of La Plata and Superior Researcher of the National Research Council of Argentina. A former President of the Argentine Astronomical Society, he has published more than 350 papers on astrophysics, gravitation, and the foundation of physics, and 10 books. His main current interest is on black hole physics and ontological problems of spacetime theories.   

\section*{Daniela P\'erez and Federico L\'opez Armengol} Fellows of CONICET under the supervision of Prof. Romero. Dr. P\'erez is a postdoctoral researcher working on black holes. Mr. L\'opez Armengol is a graduate student researching on different aspects of gravitation for his PhD Thesis. Both have broad interests in philosophy, especially metaphysics.


\begin{thebibliography}{99}

\bibitem{Bunge1967b}
Bunge, M. (1967). \textit{Foundations of Physics}. Berlin, Heidelberg, and New York: Springer-Verlag.

\bibitem{clarke}
Clarke, C.J.S. (1977). Time in General Relativity. In J.S. Earman, C. N. Glymour, \& J.J. Stachel (Eds.), \textit{Foundations of Space-Time Theories. Minnesota Studies in the Philosophy of Science, 8} (pp. 94-108). Minnesota: University of Minnesota Press.

\bibitem{Faraoni-libro}
Faraoni, V. (2015). \textit{Cosmological and Black Hole Apparent Horizons}. Lectures Notes in Physics. Switzerland:  Springer International Publishing. 

\bibitem{Gao}
Gao, C., Chen, X., Faraoni, V., Shen Y. (2008). Does the mass of a black hole decrease due to the accretion of phanton energy? \textit{Physical Review D} 78, 024008 

\bibitem{Gold}
Gold, T. (1962). The arrow of time.  \textit{American  Journal of  Physics} 30(6), 403-410. 

\bibitem{Hawking}
Hawking, S.W. (1968). Gravitational radiation in an expanding universe. \textit{Journal of Mathematical Physics} 9, 598-604.  

\bibitem{Hawking2}
Hawking, S.W. (1974). Black hole explosions. \textit{Nature} 248, 30-31. 

\bibitem{Hawking3}
Hawking, S.W. (1975). Particle creation by black holes.  \textit{Communications in Mathematical Physics} 43, 199-220.

\bibitem{Hawking1}
Hawking, S.W. (1985). Arrow of time in cosmology.  \textit{Physical Review D} 32, 2489-2495.  

\bibitem{Hayward}
Hayward, S.A. (1994). Quasilocal gravitational energy.  \textit{Physical Review D} 49, 831-839. 

\bibitem{McVittie}
McVittie, G.C. (1933). The mass-particle in an expanding universe. \textit{Monthly Notices of the Royal Astronomical Society} 93, 325-339. 

\bibitem{Nandra1}
Nandra, R., Lasembly, A.N., Hobson, M.P. (2012). The effect of a massive object on an expanding universe.  \textit{Monthly Notices of the Royal Astronomical Society} 422, 2931-2944.

\bibitem{Nandra2}
Nandra, R., Lasembly, A.N., Hobson, M.P.  (2012). The effect of an expanding universe on massive objects. \textit{Monthly Notices of the Royal Astronomical Society} 422, 2945-2959.

\bibitem{Nolan98}
Nolan, B.C. (1998). A point mass in an isotropic universe. Existence, uniqueness and basic properties. \textit{Physical Review D} 58, 064006. 


\bibitem{Penrose1979}
Penrose, R. (1979). Singularities and time-asymmetry. In S.W. Hawking, W. Israel (Eds.), \textit{General Relativity: An Einstein Centennial} (pp. 581-638). Cambridge: 
Cambridge University Press.

\bibitem{Permu}
Perlmutter S. et al. (1999). Measurements of $\Omega$ and $\Lambda$ from 42 High-Redshift Supernovae. \textit{Astrophysical Journal}  517, 565-586. 

\bibitem{PlanckCo}
Planck Collaboration 2014. Planck 2013 results. XVI. Cosmological parameters. \textit{  Astronomy \& Astrophysics} 571, id.A16, 66 pp.

\bibitem{mention}
Romero, G.E., \&  P\'erez, D. (2011). Time and irreversibility in an accelerating universe. \textit{International Journal of Modern Physics D} 20, 14, 2831-2838. 

\bibitem{Romero-philo}
Romero, G. E. (2014). Philosophical Issues of Black Holes. In A. Barton (Ed.), \textit{Advances in Black Holes Research} (pp. 27-58). New York:
Nova Science Publishers.

\bibitem{RomeroVila2013}
Romero, G.E., \& Vila, G.S. (2014). \textit{Introduction to Black Hole Astrophysics}. Lectures Notes in Physics. Berlin: Springer. 

\bibitem{sciama}
Sciama,D. (1967). Retarded potentials and the expansion of the universe. In T. Gold, \& D.L. Schumacher (Eds.), \textit{The Nature of Time} (pp. 55-67). Ithaca: Cornell University Press.

\bibitem{wald}
Wald, R. M. (1984). \textit{General Relativity}. Chicago: Chicago University Press.

\bibitem{Wheeler1}
Wheeler, J.A., \& Feynman, R.P. (1945). Interaction with the Absorber as the Mechanism of Radiation. \textit{Reviews of Modern Physics} 17, 157-181. 

\bibitem{Wheeler2}
Wheeler, J.A., \& Feynman, R.P. (1949). Classical Electrodynamics in Terms of Direct Interparticle Action. \textit{Reviews of Modern Physics} 21, 425-433.  

\end{thebibliography}
\end{document}